\documentclass{eptcs}
\providecommand{\event}{ADG 2021} 
\usepackage{graphicx}
\usepackage{mathtools}
\usepackage{amssymb}

\newenvironment{proof}[2] {#1 #2}{\hfill$\square$}

\newcommand{\sdist}[1]{$\overline{#1}$}
\newcommand{\sarea}[1]{$\mathcal{S}_{#1}$}
\newcommand{\pdiff}[1]{$\mathcal{P}_{#1}$}
\newcommand{\amline}[2]{$\left(L\!I\!N\!E\;#1\;#2\right)$}

\newtheorem{EL}{Elimination Lemma}
\newtheorem{ACL}{Area Coordinate Lemma}

\def\orcidID#1{\unskip$^{\mbox{\href{https://orcid.org/#1}{\scriptsize{[#1]}} }}$}

\newtheorem{theorem}{Theorem}
\newtheorem{definition}{Definition}

\begin{document}

\title{The Area Method in the Wolfram Language}

\author{Jack Heimrath\orcidID{0000-0002-9334-4667}
\institute{Wolfram$\vert$Alpha \\ 
100 Trade Centre Drive \\
Champaign, IL 61820, US}
\email{jackh@wolfram.com}}

\def\titlerunning{The Area Method in the Wolfram Language}
\def\authorrunning{Heimrath}

\maketitle    

\begin{abstract}
The area method is a decision procedure for geometry developed by Chou et al. in the 1990's. The method aims to reduce the specified hypothesis to an algebraically verifiable form by applying elimination lemmas. The order in which the lemmas are applied is determined by the stated conjecture and the underlying geometric construction. In this paper we present our implementation of the area method for Euclidean geometry as a stand-alone Mathematica package.
\end{abstract}

\section{Introduction}

The area method is a semi-algebraic decision procedure for a subset of Euclidean geometry, developed by Chou, Gao, and Zhang \cite{CGZ}. Its main feature, cited by its authors as an advantage over other decision procedures for geometry, is that it generates shorter, more human-readable proofs of many theorems when compared to other methods. This is achieved by carefully keeping track of the construction steps used when setting up a particular geometric construction and by only allowing conjectures of the form $E_1=E_2$ where $E_1$ and  $E_2$ are arithmetic expressions in geometric quantities. Points occurring in the conjecture are then ``eliminated'' in the reverse order of their construction using appropriate Elimination Lemmas.

The method has been successfully implemented before, and we list the implementations we are aware of below:
\begin{itemize}
    \item the first implementation was by Chou, Gao, and Zhang;
    \item by Ye in \textit{Java} \cite{Y};
    \item by Julien Narboux in \textit{Coq} \cite{N};
    \item by Predrag Jani\v ci\'c and Pedro Quaresma in GCLC \cite{JQ};
    \item by Judit Robu in \textit{Theorema} \cite{R}.
\end{itemize}
These implementations differ in efficiency and various details, especially in the methods they use to deal with non-degeneracy conditions. Nonetheless, they are all capable of proving many non-trivial geometry theorems.

\section{The Area Method}

 In this section we give an overview of the main ideas and definitions required by the area method. For more detailed treatments see~\cite{CGZ,JNQ}.

\subsection{Geometric Quantities}\label{sec:geo_quant}

In the context of the area method a conjecture is a polynomial equation\footnote{Our implementation is also capable of dealing with inequalities.} in geometric quantities of points defined within a construction, or a logical combination of such equations or inequalities. To explain exactly what this means we must first introduce some primitive notions and definitions. First, there is the set of \textit{points} $\mathbb{P}$, which can intuitively be identified with points of the Euclidean plane. Next, there is a primitive binary function $\overline{\bullet\bullet}:\mathbb{P}^2\rightarrow\mathbb{R}$, called the \textit{signed distance}, and a primitive ternary function $\mathcal{S}_{\bullet\bullet\bullet}:\mathbb{P}^3:\rightarrow\mathbb{R}$, called the \textit{signed area}. Additionally, we introduce the very important \textit{Pythagorean difference} function, $\mathcal{P}_{\bullet\bullet\bullet}:\mathbb{P}^3\rightarrow\mathbb{R}$, defined by the formula $\mathcal{P}_{xyz}\coloneqq\overline{xy}^2+\overline{yz}^2-\overline{xz}^2$. The signed area and Pythagorean difference functions allow us to reinterpret the usual notions of ``parallel" and ``perpendicular" lines as follows: $AB\parallel CD \iff \mathcal{S}_{ACD}=\mathcal{S}_{BCD}$ and $AB\perp CD \iff \mathcal{P}_{ACD}=\mathcal{P}_{BCD}$. Finally, for notational convenience, we also define the \textit{quaternary signed area} $\mathcal{S}_{wxyz}\coloneqq\mathcal{S}_{wxy}+\mathcal{S}_{wyz}$ and \textit{quaternary Pythagorean difference} $\mathcal{P}_{wxyz}\coloneqq\mathcal{P}_{wxz}-\mathcal{P}_{yxz}$. We will talk more about these functions and various propositions governing their behaviour in Sections \ref{sec:axioms} and \ref{sec:elimination}; for now we simply state the following definition:

\begin{definition}\label{def:geometric quantity}
A \textit{geometric quantity} is any one of the following: a ratio of signed distances $\frac{\overline{ab}}{\overline{cd}}$, subject to the constraints that $\overline{cd}\neq0$ and the lines through \textit{ab} and \textit{cd} are parallel; a signed area \sarea{abc} or \sarea{abcd}; a Pythagorean difference \pdiff{abc} or \pdiff{abcd}.
\end{definition}

The most obvious shortcoming of this definition is that it renders the area method incapable of dealing with ratios of non-parallel line segments. This turns out not to be much of a problem in practice though.  Given three distinct, non-collinear points $a,b,c$ instead of considering the ill-defined geometric quantity $\frac{\overline{ab}}{\overline{bc}}$ it is often possible to restate the problem to instead consider the quantity $\frac{\mathcal{P}_{aba}}{\mathcal{P}_{bcb}}=\frac{\overline{ab}^2}{\overline{bc}^2}$, which is a valid arithmetic expression in geometric quantities.

\par
With Definition~\ref{def:geometric quantity} in hand it becomes possible to express many standard predicates of Euclidean geometry in the language of the area method. See Table~\ref{tab:translating_props} for some commonly used examples.
\begin{table}[ht]
\centering
    \caption{Example Euclidean predicates and their area method equivalents}
    \begin{tabular}{|c|c|}
        \hline
        Euclidean property & Restated with geometric quantities \\ \hline
        Points $A$ and $B$ are identical & \pdiff{ABA}=0 \\
        Points $A$, $B$, and $C$ are collinear & \sarea{ABC}=0 \\
        Line segments \sdist{AB} and \sdist{CD} are parallel & \pdiff{ABA}$\neq0\land$\pdiff{CDC}$\neq0\land$\sarea{ACD}=\sarea{BCD} \\
        Line segments \sdist{AB} and \sdist{CD} are perpendicular & \pdiff{ABA}$\neq0\land$\pdiff{CDC}$\neq0\land$\pdiff{ACD}=\pdiff{BCD} \\ \hline
    \end{tabular}
    \label{tab:translating_props}
\end{table}

\subsection{Constructions}\label{sec:constructions}

In total, there are five Elementary Construction Steps (ECS's), denoted ECS1, ECS2, ECS3, ECS4, and ECS5, utilized by the area method. We give a brief description of each of them below. In what follows, we let \amline{U}{V} denote the unique line through the points $U$ and $V$:
\begin{enumerate}
    \item construct an unconstrained point $Y$, denoted $ECS1(Y)$. This is the only ECS which can be invoked without first defining any other points, and as such is used to initialize any and all geometric constructions. Points introduced by this step are called \textit{free points}. In practice it is often convenient to introduce many free points at a time --- for this purpose we introduce the notation $ECS1(Y_1,...,Y_n)$ which is to be understood as $n$ consecutive applications of ECS1;
    \item constructs a point $Y$ such that it is the intersection of \amline{U}{V} and \amline{P}{Q}, denoted by $ECS2(Y,U,V,P,Q)$;
    \item constructs a point $Y$ such that it is the foot from a given point $P$ to \amline{U}{V}, denoted $ECS3(Y,P,$ $U,V)$;
    \item  constructs a point $Y$ on the line passing through the point $W$ and parallel to \amline{U}{V} such that $\frac{\overline{WY}}{\overline{UV}}=r$, denoted $ECS4(Y,W,U,V,r)$. Note that $r$ can be a real number, a geometric quantity, or a variable;
    \item constructs a point $Y$ on the line passing through the point $U$ and perpendicular to \amline{U}{V} such that $4\mathcal{S}_{UVY}=r\mathcal{P}_{UVU}$, denoted $ECS5(Y,U,V,r)$. Note that $r$ can be a real number, a geometric quantity, or a variable.
\end{enumerate}

Each ECS introduces exactly one new point to the construction per application. For each $1\leq i\leq l$ the points used by the step $C_i$ must already be introduced by some construction steps appearing earlier in the construction $\mathcal{C}$. The point introduced by step $C_i$ is said to be of order \textit{i} in $\mathcal{C}$, or simply to have order \textit{i} if there is no confusion about which construction is being considered.

\par
For a construction step to be well-defined certain conditions, called non-degeneracy conditions (ndg's), may be required\footnote{For a list of all necessary ndg's see \cite{JNQ}.}. Some ECS's also require a parameter to be provided as an argument --- this may be a real number or an unevaluated parameter \textit{r}. Combined, these steps can be used to reproduce a large subset of classical straightedge and compass constructions. In practice we found that the classical constructions which \textit{cannot} be recreated are the ones which involve arbitrary intersections of circles. On the flip side, some non-classical constructions are also possible thanks to the fact that the parameter \textit{r} can be any real number. This makes it possible to, for instance, square the circle.

We can now meaningfully state the following definition.
\begin{definition}
A geometric construction is a finite list $\mathcal{C}=\left(C_1,C_2,...,C_l\right)$ where, for each $1\leq i\leq l$, $C_i$ is an ECS.
\end{definition}

\subsection{Axioms}\label{sec:axioms}

The area method implicitly assumes the existence of a base field $\mathbb{F}$, where $char\left(\mathbb{F}\right)\neq2$, together with all the usual field axioms. In general, this field serves as the domain for the signed distance and signed area functions. For the purposes of our implementation, and throughout the remainder of this paper, we set $\mathbb{F}=\mathbb{R}$.

The axiomatization we chose is the one described in \cite{JNQ}, which itself is a modification of the original axiomatization proposed by Chou, Gao, and Zhang. It consists of the following universally quantified axioms:
\begin{enumerate}
    \item \sdist{AB} $=0\iff A=B$
    \item \sarea{ABC} = \sarea{CAB}
    \item \sarea{ABC} = -\sarea{ACB}
    \item \sarea{ABC} $=0 \implies$ \sdist{AB} + \sdist{BC} = \sdist{AC}
    \item $\exists\,A,B,C\in\mathbb{P}\,s.t.\,\mathcal{S}_{ABC}\neq 0$ (not all points are collinear)
    \item $\mathcal{S}_{ABC}=\mathcal{S}_{ABD}+\mathcal{S}_{ADC}+\mathcal{S}_{DBC}$ (all points lie in the same plane)
    \item $\forall\,r\in\mathbb{R}\,\exists\,P\in\mathbb{P}\,s.t.\,\mathcal{S}_{ABP}=0$ and \sdist{AP}$\,=r$\sdist{AB} (there exists a line though any two points)
    \item if $A\neq B,\,\mathcal{S}_{ABP}=0,\,\overline{AP}=r\overline{AB},\,\mathcal{S}_{ABP'}=0\,$and$\,\overline{AP'}=r\overline{AB}$ then $P=P'$ (any~two points define a unique line)
    \item if $PQ\parallel CD$ and $\frac{\overline{PQ}}{\overline{CD}}=1$ then $DQ\parallel PC$
    \item if $\mathcal{S}_{PAC}\neq0$ and $\mathcal{S}_{ABC}=0$ then $\frac{\overline{AB}}{\overline{AC}}=\frac{\mathcal{S}_{PAB}}{\mathcal{S}_{PAC}}$
    \item if $C\neq D$ and $AB\perp CD$ and $EF\perp CD$ then $AB\parallel EF$
    \item if $A\neq B$ and $AB\perp CD$ and $AB\parallel EF$ then $EF\perp CD$
    \item if $FA\perp BC$ and $\mathcal{S}_{FBC}=0$ then $4\mathcal{S}_{ABC}^2=\overline{AF}^2\overline{BC}^2$ (area of a triangle)
\end{enumerate}

\subsection{Lemmas}\label{sec:elimination}

The various lemmas used by the area method can broadly be divided into three categories: basic propositions, Elimination Lemmas, and Area Coordinate Lemmas. We give a brief overview of each category below.

\par
Basic propositions are simple consequences of the axioms described in the previous section. Combined, they guarantee the existence of a canonical form for each possible geometric quantity. This makes it possible to apply the appropriate Elimination Lemmas to a given conjecture. They are Lemmas 2.1-2.8 of \cite{JNQ}.

\par
The Elimination Lemmas are the main workhorses of the area method. Each of the lemmas expresses a geometric quantity involving a constructed point $Y$ as a rational expression of geometric quantities not involving $Y$. See Appendix~\ref{sec:elimlemmas} for the full statements of these lemmas. By applying them in the appropriate, i.e. determined by the construction, order they make it possible to eliminate all occurrences of constructed points from the conjecture. This transforms the conjecture $E_1=E_2$ into an equality of polynomials $E_1'=E_2'$, involving only free points, which in most cases can be algebraically verified. For a full description of the properties and structure of these lemmas see Section 2.4 of \cite{JNQ}.

\par
In some cases the transformed conjecture $E_1'=E_2'$ may still contain dependant variables. By artificially introducing an orthonormal basis into the construction it is possible to further reduce the conjecture by applying the so called Area Coordinate Lemmas. See Appendix~\ref{sec:elimlemmas} for the full statements of these lemmas and the \textit{Free Points and Area Coordinates} Section of \cite{JQ2} for a detailed explanation of how they work.

\subsection{An Example Proof}\label{sec:example_1}
We finish this section by giving a proof-sketch of the intercept theorem using the area method. We will also revisit this example in Section \ref{sec:implementation}, after we discuss our implementation in more detail.

\begin{theorem}[intercept theorem]\label{th:intercept}
Consider 4 distinct lines $l, m, n, o$ such that the lines $l, m$ coincide at a point $S$, the lines $n, o$ are parallel, and $S\notin n, o$. Let $A$ be intersection of lines $l, n$, $B$ the intersection of $l, o$, $C$ the intersection of $m, n$, and $D$ the intersection of $m, o$. Then the following identity holds:
\begin{equation}\label{eq:intercept}
    \frac{\overline{SA}}{\overline{AB}}=\frac{\overline{SC}}{\overline{CD}}.
\end{equation}
\end{theorem}

One possible way to set up this construction proceeds as follows:
\begin{itemize}
    \item use ECS1 three times to introduce the free points $A$, $B$, and $C$;
    \item use ECS4 to construct an arbitrary point $D$ on the line through $B$ and parallel to the line through $A$ and $C$;
    \item use ECS2 to construct the point $S$ as the intersection of the line through $A$ and $B$ with the line through $C$ and $D$.
\end{itemize}
We can write this more formally as:
\begin{equation}\label{eq:intercept_construction}
    \mathcal{C}_1=\left(ECS1(A,B,C),ECS4(D,B,A,C,r),ECS2(S,A,B,C,D)\right),
\end{equation}
where \textit{r} is an unconstrained, real parameter.

\begin{proof}[Proof sketch.]
The point $S$ was the last point added to $\mathcal{C}_1$, so it must be the first point eliminated from Equation~\ref{eq:intercept}. Using the appropriate Elimination Lemmas we find that:
\begin{equation}
    \begin{split}
        \frac{\overline{SA}}{\overline{AB}}&=\frac{\mathcal{S}_{ACD}}{\mathcal{S}_{ABC}-\mathcal{S}_{ABD}}\\
        \frac{\overline{SC}}{\overline{CD}}&=\frac{\mathcal{S}_{ABC}}{\mathcal{S}_{ACD}-\mathcal{S}_{BCD}}.
    \end{split}
\end{equation}
Substituting this into Equation~\ref{eq:intercept} we find that:
\begin{equation}\label{eq:intercept_1}
    \frac{\mathcal{S}_{ACD}}{\mathcal{S}_{ABC}-\mathcal{S}_{ABD}}=\frac{\mathcal{S}_{ABC}}{\mathcal{S}_{ACD}-\mathcal{S}_{BCD}}.
\end{equation}
Equation~\ref{eq:intercept_1} is an equality of two rational expressions in the points $A,B,C,D$ --- crucially, it no longer contains the point $S$. Next, we wish to eliminate all occurrences of the point $D$. Invoking the appropriate Elimination Lemmas once more we find the following relations:
\begin{equation}
    \begin{split}
        \mathcal{S}_{ACD}&=-\mathcal{S}_{ABC}\\
        \mathcal{S}_{BCD}&=-r\,\mathcal{S}_{ABC}\\
        \mathcal{S}_{ABD}&=r\,\mathcal{S}_{ABC},
    \end{split}
\end{equation}
where \textit{r} is the parameter introduced by ECS4 in \ref{eq:intercept_construction}. Making the substitutions gives:
\begin{equation}\label{eq:intercept_2}
    \frac{-\mathcal{S}_{ABC}}{\mathcal{S}_{ABC}-r\mathcal{S}_{ABC}}=
    \frac{\mathcal{S}_{ABC}}{-\mathcal{S}_{ABC}+r\mathcal{S}_{ABC}}
\end{equation}

and simplifying reduces to the obvious equality $1=1$, which completes the proof.
\end{proof}

Note that this proof fails if $r=1$. In that case the denominators on both sides of Equality~\ref{eq:intercept_2} are equal to 0. However, if this were the case, the line through the point $A$ and $B$ would be parallel to the line through the points $C$ and $D$, meaning that point $S$ of $\mathcal{C}_1$ would be ill-defined, and hence Equation~\ref{eq:intercept} would be ill-posed. The proof fails for a similar reason if $\mathcal{S}_{ABC}=0$, as this would result in a failure of the geometric construction.

\section{The Area Method in \textit{Mathematica}}\label{sec:implementation}

Broadly speaking, our implementation consists of two parts:
\begin{itemize}
    \item the geometric framework, by which we mean both geometric constructions and geometric quantities;
    \item the area method algorithm, which includes both the elimination algorithm as well as the area coordinates algorithm;
\end{itemize}
Note that in this section we make use of \textit{Wolfram Language} specific terminology. To avoid ambiguity, we will capitalize and italicize these terms.

\subsection{The Geometric Framework}\label{sec:geo_framework}

\subsubsection{Geometric Quantities}

The signed distance, signed area, and Pythagorean difference functions are implemented simply as \textit{Symbols} with no rules for evaluation for general arguments. We found that not evaluating the Pythagorean difference immediately according to its definition as given in Section~\ref{sec:geo_quant} made it easier to apply certain Elimination Lemmas. In our experience this had no affect on the method's ability to prove a theorem. We only apply the definition of the Pythagorean difference after all possible Elimination Lemmas have already been applied (but before an optional transition to area coordinates occurs).

\subsubsection{Defining Constructions}

Our implementation of geometric constructions is very simple --- each construction is an \textit{Association}\footnote{\textit{Associations} are symbolically indexed lists.}, whose \textit{Keys} are points introduced by ECS's and whose \textit{Values} are \textit{Associations} storing all the information about a specific point. For instance, one possible way to specify the construction in Theorem~\ref{th:intercept} is as follows:

\begin{verbatim}
Input:
ECS2[s,a,b,c,d]@ECS4[d,b,a,c,r]@ECS1[a,b,c]
Output: 
<|
a-> <|"ECS"->1,"points"->{},"parameters"->{},"order"->1|>,
b-> <|"ECS"->1,"points"->{},"parameters"->{},"order"->2|>,
c-> <|"ECS"->1,"points"->{},"parameters"->{},"order"->3|>,
d-> <|"ECS"->4,"points"->{b,a,c},"parameters"->{r},"order"->4|>
s-> <|"ECS"->2,"points"->{a,b,c,d},"parameters"->{},"order"->5|>
|>
\end{verbatim}
We have also implemented the construction stating predicates listed at \cite{Nweb}.

\subsubsection{Stating Conjectures}

There are two ways to state a conjecture in our package. One way is to explicitly write down a polynomial equation in geometric quantities and use it as the input for our prover. A conjecture might look as follows:

\begin{verbatim}
    SignedArea[x,y,z]-PythagoreanDifference[x,y,z]==0.
\end{verbatim}

Note that prior to specifying a construction there is no way to resolve this conjecture. However, this input method is not very practical or intuitive to someone not familiar with the area method. To make it easier to ask geometrically meaningful questions we have implemented some goal-stating predicates, e.g. collinear, perpendicular, etc.

Our prover also accepts inequalities as input.

\subsubsection{Verifying ndg's}

Recall that each ECS (with the exception of ECS1) has some associated ndg's which must be verified. Every time a construction is evaluated, we check whether the \textit{negation} of the specific ndg is a provable theorem (by means of the area method itself) within the appropriate construction. If the negation of the ndg is not provable we assume the construction to be consistent. This means that, from the point of view of our prover, provable statements and consistent constructions are only \textit{generally} provable statements and \textit{generally} consistent constructions. Henceforth, whenever we call a statement provable or a construction consistent we will mean that in the general sense. For a detailed discussion of the completeness of the method see Section~2.5.8 of~\cite{JNQ}.

Thus in the provided example we first check whether the statement $\mathcal{P}_{aca}=0$ is provable in the construction ECS1[a,b,c]. The points $a,b,c$ are free points and as such they are assumed to be generally distinct, meaning the statement $\mathcal{P}_{aca}=0$ is not generally true, and hence the appropriate ndg is satisfied. Similarly we check if $\mathcal{S}_{cab}=\mathcal{S}_{dab}$ is provable in the construction ECS4[d,b,a,c,r]@ECS1[a,b,c]. This statement is also in general not provable, and so the ndg is assumed to be satisfied. However, care must be taken: setting the previously unevaluated parameter $r$ in ECS4 to be equal to 1 and evaluating the construction again would lead to the first contradiction discussed in Section~\ref{sec:example_1} and the construction would not evaluate properly.

In the original formulation of Chou, Gao, and Zhang, ratios of signed distances were only defined for parallel line segments. In our implementation we make no such restrictions, and a conjecture of the form:
\begin{equation}\label{eq:alg_verif}
    \frac{\overline{AB}}{\overline{CD}}=\frac{\overline{AB}}{\overline{CD}}
\end{equation}
would be a valid theorem, assuming $\overline{CD}\neq0$ and the points $A,B,C,D$ are all defined, even if the line segments $AB$ and $CD$ are not parallel. This is because Equation~\ref{eq:alg_verif} is algebraically verifiable. Instead, our implementation of the Elimination Lemmas for ratios of signed distances only applies to ratios of provably parallel line segments.

Typically, the area method also requires one to check additional ndg's to ensure that the denominators of ratios of signed distances are non-zero. However, thanks to the symbolic computation capabilities of the \textit{Wolfram Language}, any occurrence of a zero in the denominator of any expression results in a ``division by zero" error message and prevents the correct evaluation of our program.

\subsubsection{Uniformization and Simplification.} Uniformization and simplification rules are consequences of the basic propositions mentioned in \ref{sec:elimination}. Uniformization of a  conjecture $E_1=E_2$ is the process of transforming each geometric quantity occurring in the equality into its canonical form. The arguments of each function are sorted according to their order (if no construction is explicitly specified the points will be sorted according to the canonical order of their names as \textit{Wolfram Language} \textit{Symbols}), and an appropriate sign change is applied if necessary to preserve orientation. We chose to implement this procedure as conditional evaluation rules for our functions, the condition being that the arguments are not sorted according to their construction order. This ensures that each geometric quantity is always in the appropriate form for an Elimination Lemma to be applied. It also  ensures that syntactically different expressions describing the same quantity, for example \sarea{ABC} and \sarea{BCA}, are provably equal.

Simplification is the process of applying field operations as well as evaluating degenerated geometric quantities. For the field operations we made do with the \textit{Wolfram Language}'s built in symbolic computation capabilities. Degenerated geometric quantities are quantities of the type \sdist{AA}, \sarea{AAB}, \pdiff{AAB}, etc. They correspond to degenerate line segments and triangles and are all identically equal to zero. This procedure is also implemented as conditional evaluation rules for the appropriate geometric quantities. It helps reduce the size and complexity of the conjecture $E_1=E_2$ during the elimination process and ensures that any ``division by zero" errors are immediately caught.

\subsection{The Area Method Algorithm}
The elimination algorithm works by applying appropriate Elimination Lemmas to the conjecture $E_1=E_2$ in an order determined by the construction. After all possible lemmas have been applied, the conjecture is transformed into a reduced form $E_1'=E_2'$. We now have the following scenarios:
\begin{enumerate}
    \item $E_1'=E_2'$ is not expressed solely in terms of geometric quantities of free points. In this case the conjecture is not a theorem and the algorithm returns the statement $E_1'=E_2'$ and terminates;
    \item $E_1'=E_2'$ is an algebraically provable polynomial equation in free points. In this case the conjecture is a theorem and the algorithm returns \textit{True} and terminates;
    \item $E_1'=E_2'$ is a polynomial equation in free points, but is not algebraically provable. In this case the area coordinates algorithm is applied to the reduced conjecture.
\end{enumerate}

\subsubsection{The Elimination Algorithm}
The elimination algorithm works by repeatedly applying the following procedure:
\begin{itemize}
    \item determine the last non-free point $Y$ occurring in the current conjecture statement;
    \item try to replace every geometric quantity containing $Y$ with an expression determined by the appropriate Elimination Lemma. The geometric quantity containing $Y$ together with the knowledge of which ECS was used to construct $Y$ are sufficient to determine which Elimination Lemma should be used;
    \item \textit{Simplify} the resulting expression, which becomes the new conjecture statement.
\end{itemize}
Once this process terminates we are left with a reduced conjecture $E_1'=E_2'$. If it is not algebraically verifiable, we expand all Pythagorean differences appearing in the conjecture according to the definition $\mathcal{P}_{abc}=\overline{ab}^2+\overline{bc}^2-\overline{ac}^2$. If after these substitutions the conjecture is still not algebraically verifiable we pass on to the area coordinates algorithm.

Some Elimination Lemmas have side conditions which alter the expression substituted into the conjecture. As was the case with ndg's, side conditions are also tested using the area method algorithm. These conditions are always mutually exclusive and exhaustive, so the first condition which can be verified determines the exact form of the Elimination Lemma. 

\subsubsection{The Area Coordinates Algorithm}\label{sec:area coordinates}
It may be the case that the elimination algorithm successfully reduces the conjecture to an arithmetic expression using only geometric quantities in free points, but which is not immediately algebraically verifiable. Heron's Formula for the area of a triangle is a good example, because the associated conjecture is stated exclusively in terms of free points. In cases such as these the area coordinate algorithm may be used to further reduce the conjecture. The first two points introduced into the construction are renamed $O$ and $X$ and a new point is added via ECS5[Y,O,X,1]. Essentially, this introduces an orthogonal basis $\{OX,OY\}$ into the construction, which W.L.O.G.we can  assume to be orthonormal. The Area Coordinate Lemmas are then applied, reducing the conjecture to an expression in the area coordinates of the free points w.r.t. the basis. These quantities are independent, making it possible to algebraically prove or disprove the conjecture.

\subsection{Example Proofs in \textit{Mathematica}}
\begin{figure}
    \centering
    \includegraphics[width=\textwidth,height=\textheight,keepaspectratio]{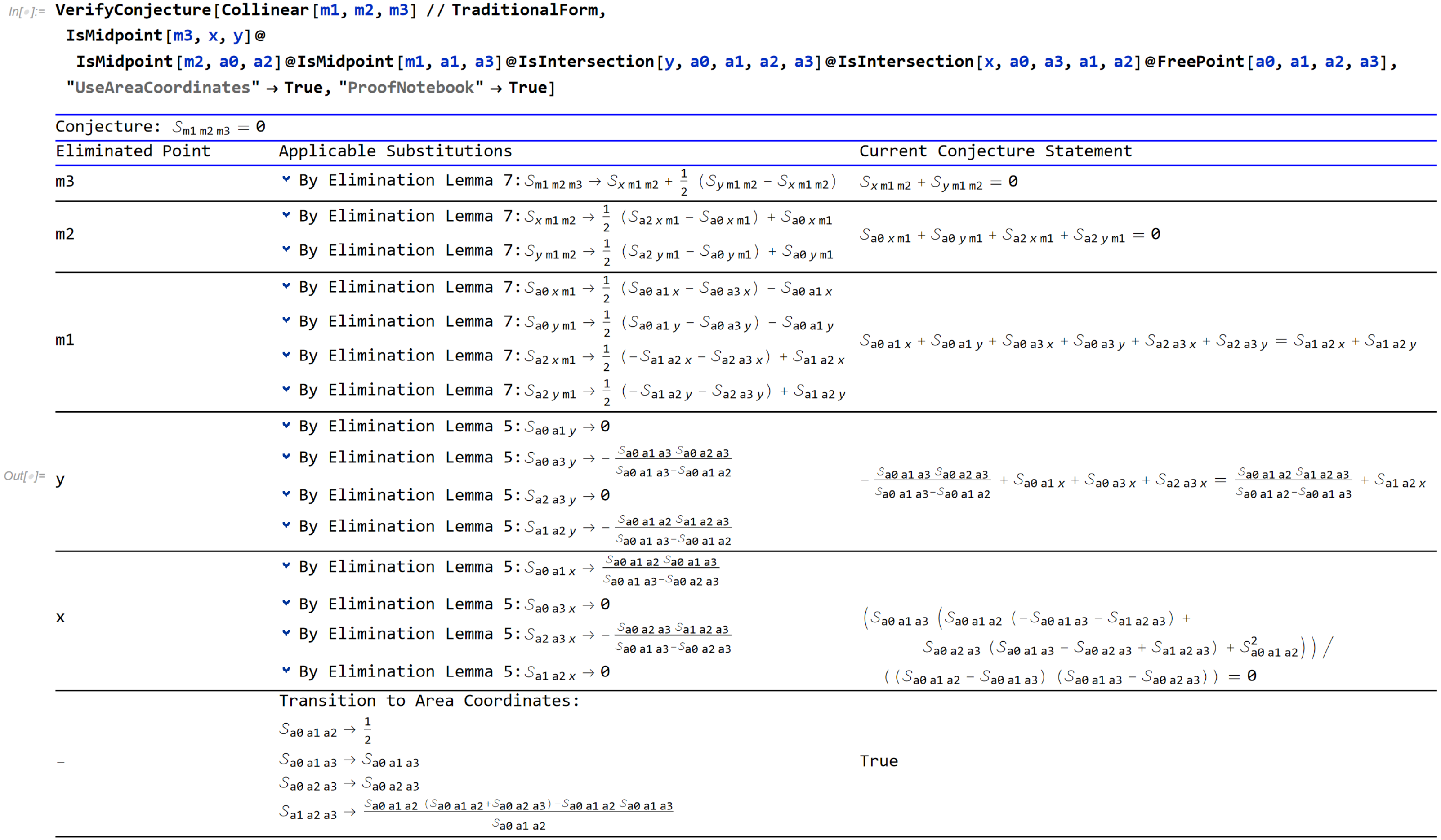}
    \caption{A full proof of the existence of the Gauss-Newton line}
    \label{fig:gaussnewton}
\end{figure}

Figure~\ref{fig:gaussnewton} contains a full proof of the existence of the Gauss-Newton line, that is the line containing all three midpoints of the diagonals of a complete quadrilateral, as generated by our prover.

\begin{figure}
    \centering
    \includegraphics[width=\textwidth,height=\textheight,keepaspectratio]{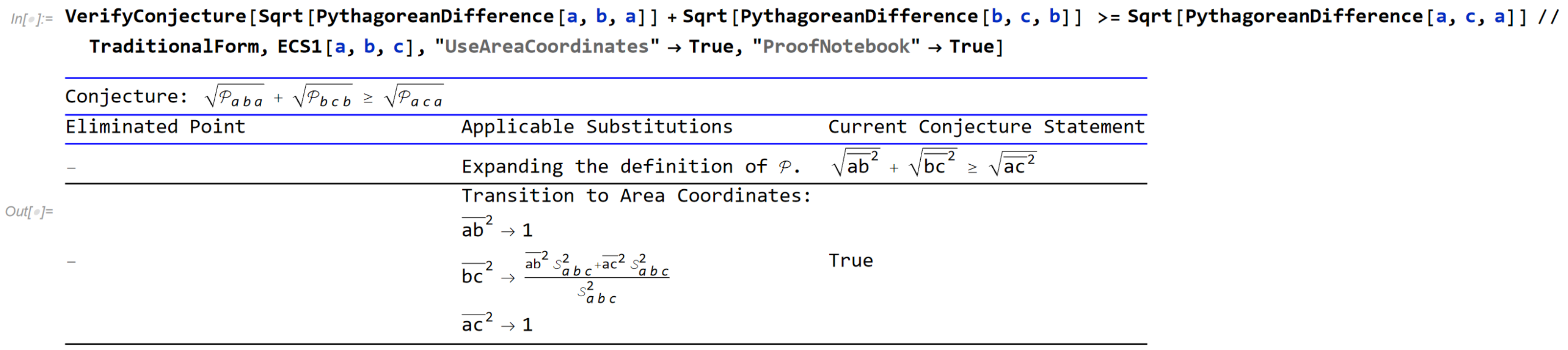}
    \caption{A proof of the triangle inequality}
    \label{fig:triangle}
\end{figure}

Figure~\ref{fig:triangle} contains a proof of the triangle inequality. The conjecture is of a form not typically considered when discussing the area method as it contains square roots\footnote{No special adjustments were made to the prover to let it handle square roots.}. Despite this we chose to include it to demonstrate that our prover handles simple inequalities, to illustrate the proof of a theorem which doesn't contain any constructed points, and finally due to the triangle inequality's importance in mathematics in general.

Finally, Table~\ref{timings} shows how long it takes to prove some famous geometric theorems. All computations were run on a fresh \textit{Mathematica} kernel, on an Intel Core i7-8650U CPU with 8GB of RAM.

\begin{table}\caption{The time (in \textit{ms}) it takes to prove some famous geometric theorems.}\label{timings}
\begin{center}
\begin{tabular}{ |c|c|c| } 
 \hline
 Theorem & Area Coordinates & Time (\textit{ms})\\  
 \hline
 Ceva's Theorem & no & 974.3 \\ 
 \hline
 Desargues Theorem & no & 495.1 \\ 
 \hline 
 Euler Line & yes & 6647.5 \\ 
 \hline 
 Gauss-Newton Line & yes & 548.9 \\ 
 \hline 
 Heron's Formula & yes & 406.4 \\ 
 \hline 
 Intercept Theorem & no & 425.7 \\ 
 \hline 
 Midpoint Theorem & no & 199.6 \\ 
 \hline 
 Menelaus' Theorem & no & 1143.7 \\ 
 \hline 
 Pappus's Line Theorem & no & 583.2 \\ 
 \hline 
 Pythagorean Theorem & no & 131.7 \\ 
 \hline 
 Triangle Inequality & yes & 219.7 \\ 
 \hline
\end{tabular}
\end{center}
\end{table}

\section{Future Work}
We see two ways of further developing the capabilities of this package. The first is to extend our implementation to higher-dimensional or non-euclidean geometries. In principle this would require only the development of a suitable axiomatization and modification of the elimination lemmas. The second is to invert the elimination procedure to search for invariants within a specific construction. While this approach could potentially require much more computational power than the standard area method, it might lead to the discovery of new theorems or simplify proofs of existing results.

Less ``theoretical" directions of future work include integrating this package into existing geometrical functionality available within the \textit{Wolfram Language} and improving it's performance in the \textit{Wolfram} cloud. This last item in particular is of note, as it makes it possible for anyone to experiment with the area method in a browser, without installing a local \textit{Wolfram} environment or buying a license.

\section{Conclusion}
In this paper we gave a description of the area method and its implementation as a stand-alone \textit{Mathematica} package. We believe this method is particularly well served by the symbolic computation and pattern matching capabilities of the \textit{Wolfram Language} --- a collection of 20 examples, including theorems of Desargues, Menelaus, and the existence of the Euler line, is proved in under 20 seconds. We have also implemented ways of dealing with all possible non-degeneracy conditions, ensuring the correctness of our results.

\subsubsection{Acknowledgements.} I would like to thank Pedro Quaresma and Predrag Jani\v ci\'c for their helpful comments during the development of the \textit{Mathematica} package and for encouraging me to write this article. I am also grateful to Bradley Janes and the \textit{Wolfram} Synthetic Geometry team for their continued support of my work and the advice they provide.

\appendix

\section{Elimination Lemmas}\label{sec:elimlemmas}
Below we provide a list of Elimination Lemmas implemented in our prover. Since the simplification of degenerate quantities precedes the application of Elimination Lemmas in our implementation we may assume that the points $A,B,C,D$ are all distinct from the point $Y$. Recall that $G(Y)$ denotes a linear quantity.

\begin{EL}
If Y is introduced by $ECS2(Y,U,V,P,Q)$ then:
\begin{equation}
    \frac{\overline{AY}}{\overline{CY}}=
    \begin{cases}
    \frac{\mathcal{S}_{APQ}}{\mathcal{S}_{CPQ}} & \text{if $A$ is on $UV$}\\
    \frac{\mathcal{S}_{AUV}}{\mathcal{S}_{CUV}} & \text{otherwise}
    \end{cases}
\end{equation}
\begin{equation}
    \frac{\overline{AY}}{\overline{CD}}=
    \begin{cases}
    \frac{\mathcal{S}_{APQ}}{\mathcal{S}_{CPDQ}} & \text{if $A$ is on $UV$}\\
    \frac{\mathcal{S}_{AUV}}{\mathcal{S}_{CUDV}} & \text{otherwise}
    \end{cases}
\end{equation}
\end{EL}

\begin{EL}
If Y is introduced by $ECS3(Y,P,U,V)$ then:
\begin{equation}
    \frac{\overline{AY}}{\overline{CY}}=
    \begin{cases}
    \frac{\mathcal{P}_{PUV}\mathcal{P}_{PCAV}+\mathcal{P}_{PVU}\mathcal{P}_{PCAU}}{\mathcal{P}_{PUV}\mathcal{P}_{CVC}+\mathcal{P}_{PVU}\mathcal{P}_{CUC}-\mathcal{P}_{PUV}\mathcal{P}_{PVU}} & \text{if $A$ is on $UV$}\\
    \frac{\mathcal{S}_{AUV}}{\mathcal{S}_{CUV}} & \text{otherwise}
    \end{cases}
\end{equation}
\begin{equation}
    \frac{\overline{AY}}{\overline{CD}}=
    \begin{cases}
    \frac{\mathcal{P}_{PCAD}}{\mathcal{P}_{CDC}} & \text{if $A$ is on $UV$}\\
    \frac{\mathcal{S}_{AUV}}{\mathcal{S}_{CUDV}} & \text{otherwise}
    \end{cases}
\end{equation}
\end{EL}

\begin{EL}
If Y is introduced by $ECS4(Y,R,P,Q,r)$ then:
\begin{equation}
    \frac{\overline{AY}}{\overline{CY}}=
    \begin{cases}
    \frac{\frac{\overline{AR}}{\overline{PQ}}+r}{\frac{\overline{CR}}{\overline{PQ}}+r} & \text{if $A$ is on $RY$}\\
    \frac{\mathcal{S}_{APRQ}}{\mathcal{S}_{CPRQ}} & \text{otherwise}
    \end{cases}
\end{equation}
\begin{equation}
    \frac{\overline{AY}}{\overline{CD}}=
    \begin{cases}
     \frac{\frac{\overline{AR}}{\overline{PQ}}+r}{\frac{\overline{CD}}{\overline{PQ}}} & \text{if $A$ is on $RY$}\\
    \frac{\mathcal{S}_{APRQ}}{\mathcal{S}_{CPDQ}} & \text{otherwise}
    \end{cases}
\end{equation}
\end{EL}

\begin{EL}
If Y is introduced by $ECS5(Y,P,Q,r)$ then:
\begin{equation}
    \frac{\overline{AY}}{\overline{CY}}=
    \begin{cases}
    \frac{\mathcal{S}_{APQ}-\frac{r}{4}\mathcal{P}_{PQP}}{\mathcal{S}_{CPQ}-\frac{r}{4}\mathcal{P}_{PQP}} & \text{if $A$ is on $PY$}\\
    \frac{\mathcal{P}_{APQ}}{\mathcal{P}_{CPQ}} & \text{otherwise}
    \end{cases}
\end{equation}
\begin{equation}
    \frac{\overline{AY}}{\overline{CD}}=
    \begin{cases}
     \frac{\mathcal{S}_{APQ}-\frac{r}{4}\mathcal{P}_{PQP}}{\mathcal{S}_{CPDQ}} & \text{if $A$ is on $PY$}\\
    \frac{\mathcal{P}_{APQ}}{\mathcal{P}_{CPDQ}} & \text{otherwise}
    \end{cases}
\end{equation}
\end{EL}

\begin{EL}
If Y is introduced by $ECS2(Y,U,V,P,Q)$ then:
\begin{equation}
    G(Y)=\frac{\mathcal{S}_{UPQ}G(V)-\mathcal{S}_{VPQ}G(U)}{\mathcal{S}_{UPVQ}}
\end{equation}
\end{EL}

\begin{EL}
If Y is introduced by $ECS3(Y,P,U,V)$ then:
\begin{equation}
    G(Y)=\frac{\mathcal{P}_{PUV}G(V)+\mathcal{P}_{PVU}G(U)}{\mathcal{P}_{UVU}}
\end{equation}
\end{EL}

\begin{EL}
If Y is introduced by $ECS4(Y,W,U,V,r)$ then:
\begin{equation}
    G(Y)=G(W)+r(G(V)-G(U))
\end{equation}
\end{EL}

\begin{EL}
If Y is introduced by $ECS5(Y,P,Q,r)$ then:
\begin{equation}
    \mathcal{S}_{ABY}=\mathcal{S}_{ABP}-\frac{r}{4}\mathcal{P}_{PAQB}
\end{equation}
\end{EL}

\begin{EL}
If Y is introduced by $ECS5(Y,P,Q,r)$ then:
\begin{equation}
    \mathcal{P}_{ABY}=\mathcal{P}_{ABP}-4r\mathcal{S}_{PAQB}
\end{equation}
\end{EL}

\begin{EL}
If Y is introduced by $ECS2(Y,U,V,P,Q)$ then:
\begin{equation}
    \mathcal{P}_{AYB}=\frac{\mathcal{S}_{UPQ}}{\mathcal{S}_{UPVQ}}G(V)+\frac{\mathcal{S}_{VPQ}}{\mathcal{S}_{UPVQ}}G(U)-\frac{\mathcal{S}_{UPQ}\mathcal{S}_{VPQ}\mathcal{P}_{UVU}}{\mathcal{S}_{UPVQ}^2}
\end{equation}
\end{EL}

\begin{EL}
If Y is introduced by $ECS3(Y,P,U,V)$ then:
\begin{equation}
    \mathcal{P}_{AYB}=\frac{\mathcal{P}_{PUV}}{\mathcal{P}_{UVU}}G(V)+\frac{\mathcal{P}_{PVU}}{\mathcal{P}_{UVU}}G(U)-\frac{\mathcal{P}_{PUV}\mathcal{P}_{PVU}}{\mathcal{P}_{UVU}}
\end{equation}
\end{EL}

\begin{EL}
If Y is introduced by $ECS4(Y,W,U,V,r)$ then:
\begin{equation}
    \mathcal{P}_{AYB}=\mathcal{P}_{AWB}+r(\mathcal{P}_{AVB}-\mathcal{P}_{AUB}+2\mathcal{P}_{WUV})-r(1-r)\mathcal{P}_{UVU}
\end{equation}
\end{EL}

\begin{EL}
If Y is introduced by $ECS5(Y,P,Q,r)$ then:
\begin{equation}
    \mathcal{P}_{AYB}=\mathcal{P}_{APB}+r^2\mathcal{P}_{PQP}-4r(\mathcal{S}_{APQ}+\mathcal{S}_{BPQ})
\end{equation}
\end{EL}

The final three lemmas are applied only if the use of area coordinates is necessary and requested by the user and if the conjecture has been reduced to an expression involving only free points. In what follows the points $O,X,Y$ are as discussed in Section~\ref{sec:area coordinates}.

\begin{ACL}
Let A, B, C be any free points, then:
\begin{equation}
    \mathcal{S}_{ABC}=\frac{(\mathcal{S}_{OYB}-\mathcal{S}_{OYC})\mathcal{S}_{OXA}+(\mathcal{S}_{OYC}-\mathcal{S}_{OYA})\mathcal{S}_{OXB}+(\mathcal{S}_{OYA}-\mathcal{S}_{OYB})\mathcal{S}_{OXC}}{\mathcal{S}_{OXY}}
\end{equation}
\end{ACL}

\begin{ACL}
Let A, B, C be any free points, then:
\begin{equation}
    \overline{AB}^2=\frac{\overline{OX}^2(\mathcal{S}_{OYA}-\mathcal{S}_{OYB})^2+\overline{OY}^2(\mathcal{S}_{OXA}-\mathcal{S}_{OXB})^2}{\mathcal{S}_{OXY}^2}
\end{equation}
\end{ACL}

\begin{ACL}
\begin{equation}
    \mathcal{S}_{OXY}^2=\frac{1}{4}
\end{equation}
\end{ACL}

\bibliography{eptcs}

\end{document}